\documentclass[superscriptaddress, showpacs,showkeys,twocolumn,showpacs,amsmath,amssymb]{revtex4}
\usepackage{graphicx}
\usepackage{dcolumn}
\usepackage{bm}
\usepackage{hyperref}
\usepackage{commath}

\usepackage{amsmath}
\usepackage{verbatim}

\begin{document}
\title{Scaling of transmission capacities in coarse-grained renewable electricity networks}





\author{Mirko Sch\"{a}fer}
\affiliation{%
Department of Engineering, Aarhus University, Inge Lehmanns Gade 10, 8000 Aarhus C, Denmark
}
\author{Simon Bugge Siggaard}
\affiliation{%
Department of Physics and Astronomy, Aarhus University, Ny Munkegade 120, 8000 Aarhus C, Denmark
}
\author{Kun Zhu}
\affiliation{%
Department of Engineering, Aarhus University, Inge Lehmanns Gade 10, 8000 Aarhus C, Denmark
}
\author{Chris Risager Poulsen}
\affiliation{%
Department of Physics and Astronomy, Aarhus University, Ny Munkegade 120, 8000 Aarhus C, Denmark
}
\author{Martin Greiner}
\affiliation{%
Department of Engineering, Aarhus University, Inge Lehmanns Gade 10, 8000 Aarhus C, Denmark
}

 	
\begin{abstract}
Network models of large-scale electricity systems feature only a limited spatial resolution, either due to lack of data or in order to reduce the complexity of the problem with respect to numerical calculations. In such cases, both the network topology, the load and the generation patterns below a given spatial scale are aggregated into representative nodes. This coarse-graining affects power flows and thus the resulting transmission needs of the system. We derive analytical scaling laws for measures of network transmission capacity and cost in coarse-grained renewable electricity networks. For the cost measure only a very weak scaling with the spatial resolution of the system is found. The analytical results are shown to describe the scaling of the transmission infrastructure measures for a simplified, but data-driven and spatially detailed model of the European electricity system with a high share of fluctuating renewable generation.
\end{abstract}

\maketitle

\section{Introduction}
\begin{figure}
\includegraphics[scale=0.37]{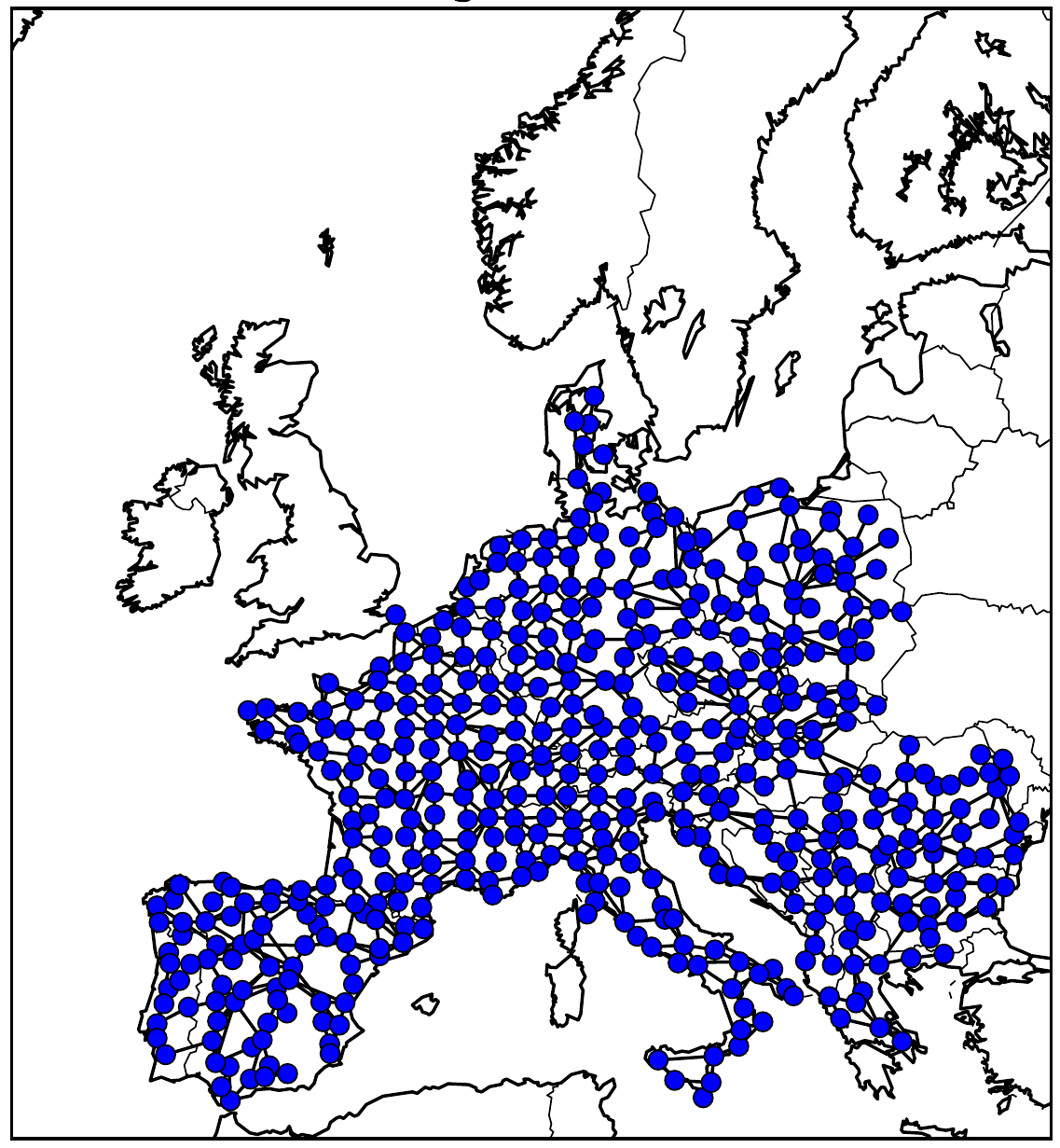}
\includegraphics[scale=0.37]{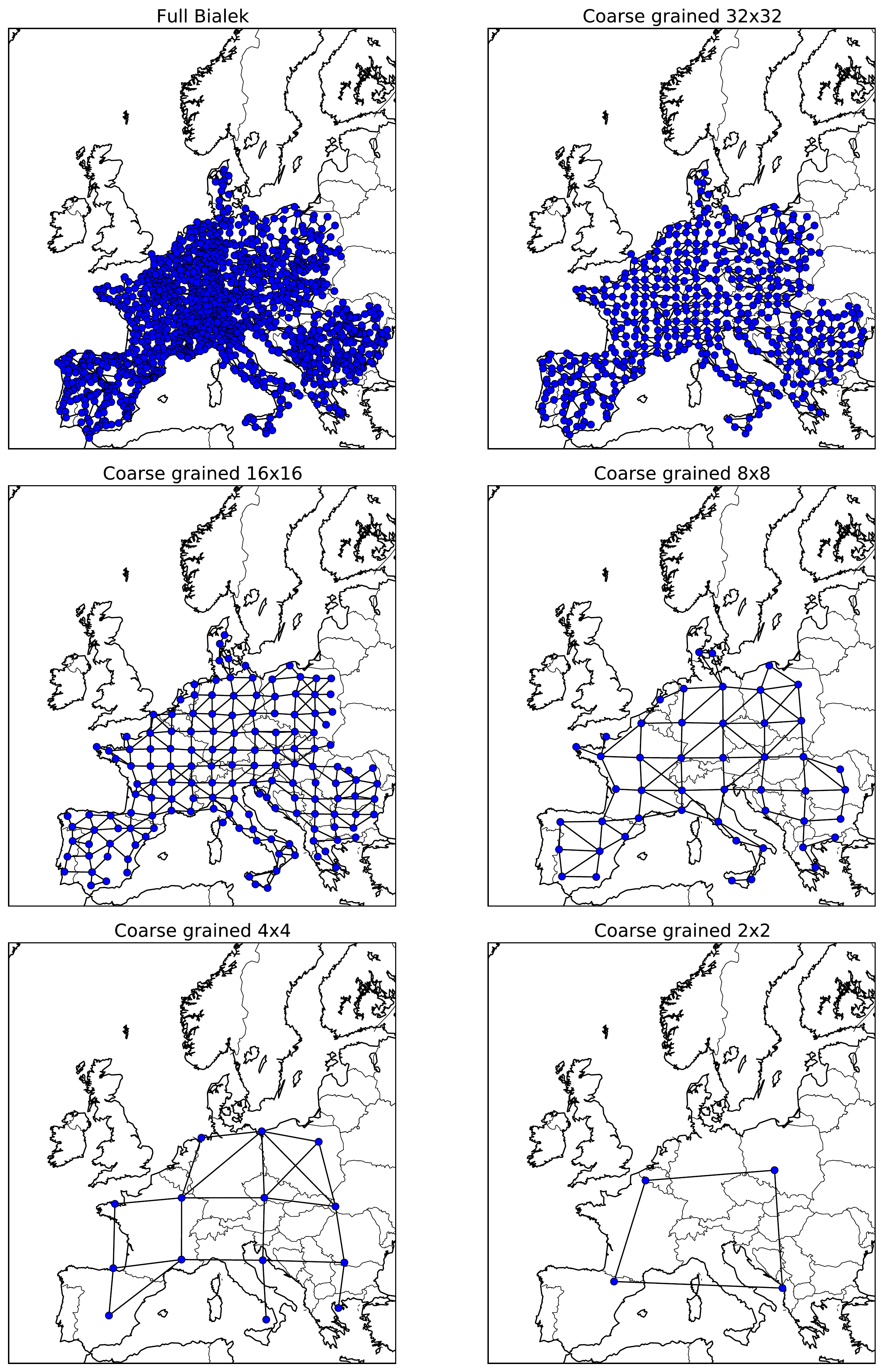}
\caption{Two exemplary clustering representations of the European transmission grid based on  $32\times 32$ grid cells (left) and $8\times 8$ (right) grid cells, respectively. 
}
\label{fig:eu_clustering}
\end{figure}

Data-driven numerical models provide important guidelines for the design of a future sustainable energy system, which presumably will depend on renewable power generation from wind and solar~\cite{sims2011}. Given the spatial heterogeneity of weather patterns, the efficient exploitation of locations with favorable renewable resource quality calls for a high spatial resolution in the respective models. The same holds for the placement of transmission infrastructure, which should be adapted to the given spatial distribution of generation and load and resolve potential bottlenecks. Unfortunately, such a detailed spatial model resolution comes with challenging demands in data quality and computational complexity. Consequently, depending on the focus and the complexity of the model one has to apply a spatial coarse-graining, which aggregates network infrastructure, load and generation patterns over spatial scales ranging from a few kilometers to entire countries~\cite{hoersch2017}. By changing the network topology as well as the spatio-temporal pattern of nodal power injections, this coarse-graining procedure has in particular an impact on the power flows and the resulting transmission infrastructure needs proposed by the model.\\
In this contribution we study for the first time the scaling of transmission properties of electricity system models under spatial clustering from a complex networks perspective. We choose a model of the European system, which combines a data-driven approach under a high spatial resolution with simplified dispatch schemes. Applying a straightforward spatial clustering algorithm (see fig.~\ref{fig:eu_clustering}), the total transmission capacity and cost of the network are evaluated dependent on the power flow statistics for different spatial resolutions. We show that in particular the total transmission capacity cost only scales weakly with the spatial resolution. To provide analytical insights, we perform various approximations for the infrastructure objectives and derive general scaling laws, which describe the numerical results remarkably accurately.\\
The article is structured as follows. We first introduce the data-driven network model of a highly renewable European electricity system. Subsequently we define the spatial coarse-graining algorithm and present numerically results for the scaling of the total transmission capacity and cost. In the subsequent section we apply a series of approximations for these measures, yielding a simplified form which allows to derive an analytical description of the scaling behaviour. A conclusion and outlook is given in the last section.
\section{A data-driven network model of a highly renewable European electricity system}
We study the scaling of transmission infrastructure measures using a simplified, but spatially detailed large-scale model of a highly renewable European electricity system~\cite{jensen2017}. The network topology representing the main European transmission grid is adapted from~\cite{hutcheon2013}, comprising $N=1494$ nodes (buses) and $L=2156$ links (transmission lines). Based on data from~\cite{jensen2017}, we construct two time series $G_{n}^{R}(t)=G_{n}^{W}(t)+G_{n}^{S}(t)$ and $L_{n}(t)$, constituting the renewable generation from wind and solar power generation, and the load in the geographical region represented by node $n$  for the three years $2012-2014$ with hourly resolution. Using a renewable energy atlas~\cite{andresen2015}, the renewable generation in this data set is obtained by converting weather data into raw generation data for solar PV and wind power generation $\tilde{G}_{n}^{S}(t)$ and $\tilde{G}_{n}^{W}(t)$, respectively~\cite{jensen2017}. The demand side in the data set is represented by the load time series $L_{n}(t)$, which is based on regionalised historical load data taken from ENTSO-E~\cite{jensen2017}. We scale the raw generation data as follows:
\begin{align}
\label{eq:layout1}
G_{n}^{W}(t) &= \tilde{G}_{n}^{W}(t) \gamma_{j}\alpha_{j} \frac{\sum_{m\in S_j}\langle L_{m} \rangle}{\sum_{m \in S_{j}}\langle \tilde{G}_{m}^{W}\rangle}~,\\
\label{eq:layout2}
G_{n}^{S}(t) &= \tilde{G}_{n}^{S}(t) \gamma_{j}(1-\alpha_{j}) \frac{\sum_{m\in S_j}\langle L_{m} \rangle}{\sum_{m \in S_{j}}\langle \tilde{G}_{m}^{S}\rangle}~.
\end{align}
Here $n$ denotes the respective node, and $S_j$ represents the set of nodes in country $j$. We choose a renewable penetration $\gamma_{j}=1$ and wind share $\alpha_{j}=0.8$ for all countries. This highly renewable layout assures that for each country \emph{on average} $100\%$~of the load is covered by renewable generation, with a mix of $80\%$ wind power and $20\%$ solar power~\cite{rodriguez2015cost}. Inside the countries, the heterogeneous distribution of renewable generation capacity according to~eqs.~(\ref{eq:layout1}) and~(\ref{eq:layout2}) makes use of favorable locations.\\
In general there will be an instantaneous local mismatch or residual load $\Delta_{n}(t) = G_{n}^{R}(t) - L_{n}(t)$, which has to be balanced by imports/exports $P_{n}(t)$ through the transmission grid, or by local generic backup power generation or curtailment $B_{n}(t)$. Here we apply the following simplified synchronised balancing scheme, which dispatches backup energy or curtails excess energy proportional to the average load of the respective node~\cite{rodriguez2015localized}:
\begin{equation}
\label{eq:sync_balancing}
B_{n}(t) = \frac{\langle L_{n} \rangle}{\sum_{m}\langle L_{m} \rangle}\sum_{k}\Delta_{k}(t)~.
\end{equation}
The nodal power injection $P_{n}(t)$ is fixed by nodal energy conservation,
\begin{equation}
\label{eq:nodal_balance}
G_{n}^{R}(t) - L_{n}(t) = \Delta_{n}(t) = B_{n}(t) + P_{n}(t)~.
\end{equation}
A positive power injection $P_{n}(t)>0$ corresponds to an exporting node, whereas a negative injection $P_{n}(t)<0$ represents a net importing node. We apply the DC approximation to the full AC power flow equations, which yields a linear relationship between the injection pattern $P_{n}(t)$ and the power flow $F_{l}(t)$ on a link $l$~\cite{purchala2005}:
\begin{equation}
\label{eq:ptdf_flow}
F_{l}(t) = \sum_{n}H_{ln}P_{n}(t)~.
\end{equation}
Here $H_{ln}$ is the matrix of power transfer distribution factors (PTDF), which incorporates information about the network topology and the line susceptances~\cite{wood2012}. Assuming for simplicity unit line susceptances, the PTDF matrix can be calculated as $\mathbf{H}=\mathbf{K}^{T}\mathbf{L}^{\dagger}$, where $\mathbf{L}^{\dagger}$ denotes the Moore-Penrose pseudo inverse of the network Laplacian $\mathbf{L}$, and $\mathbf{K}^{T}$ is the transposed incidence matrix with
\begin{equation}
K^{T}_{ln}= \left\{%
\begin{array}{ll}
1 & \mbox{if link $l$ starts at node $n$}~,\\
-1 & \mbox{if link $l$ ends at node $n$}~,\\
0 & \mbox{otherwise}~.
\end{array}\right.
\end{equation}
For a balanced injection pattern $P_{n}$ with $\sum_{n}P_{n}=0$, the power flows calculated from eq.~(\ref{eq:ptdf_flow}) are invariant  under a constant gauge $H_{lm} \to H_{lm} +c_{l}$ for the PTDF matrix. We can use the degree of freedom expressed by the constants $c_{l}$ for the incorporation of the balancing scheme in eq.~(\ref{eq:sync_balancing}) into the power flow calculation,
\begin{align}
\label{eq:gauge1}
F_{l} &= \sum_{n}H_{ln}\left(\Delta_{n} - \frac{\langle L_{n} \rangle}{\sum_{m}\langle L_{m} \rangle}\sum_{k}\Delta_{k}\right)\\
\label{eq:gauge2}
&= \sum_{n}\left(H_{ln}+\hat{c}_{l}\right)\Delta_{n}~,
\end{align}
with
\begin{equation}
\label{eq:gauge_c}
\hat{c}_{l} = - \sum_{k}H_{lk}\frac{\langle L_{k}\rangle}{\sum_{m}\langle L_{m} \rangle}~.
\end{equation}
In the following we always assume that this gauge $H_{ln}\to H_{ln} + \hat{c}_{l}$ has been performed, which allows to apply the power flow calculations directly to the (unbalanced) mismatch pattern $\Delta_{n}(t)$.

We define the transmission capacity $\mathcal{K}_{l}$ of a link $l$ as the $q=0.99$ quantile of the corresponding flow distribution~$p(F_{l})$~\cite{rodriguez2014transmission}:
\begin{equation}
\label{eq:capacity}
q = 0.99 =  \int\limits_{-\mathcal{K}_{l}}^{\mathcal{K}_{l}} p(F_{l}) d F_{l}~.
\end{equation}
This approach assumes that the power flows in the model are unconstrained, with the necessary transmission capacities  determined in retrospect from the flow statistics. The extreme events excluded by this definition are assumed to be covered by emergency measures like storage or demand side management not considered in our simplified model. We define the transmission capacity to be identical for power flows in both directions of the respective link, with the total transmission capacity of the network given as the sum over all system links $\mathcal{K}=\sum_{l}\mathcal{K}_{l}$. The cost of transmission infrastructure between two nodes is expected to be proportional to both the length of the connecting link and the respective transmission capacity. As a measure for the total transmission cost $\mathcal{T}$ of the network we thus use the estimate
\begin{equation}
\label{eq:cost}
\mathcal{T} = \sum_{l} d_{l}\mathcal{K}_{l}~,
\end{equation}
with $d_{l}$ the geodesic length of link $l$. For simplicity here we do not discriminate between links representing AC and DC transmission lines.
\section{Scaling of transmission capacity measures under coarse-graining}
\begin{figure}
\includegraphics[scale=0.5]{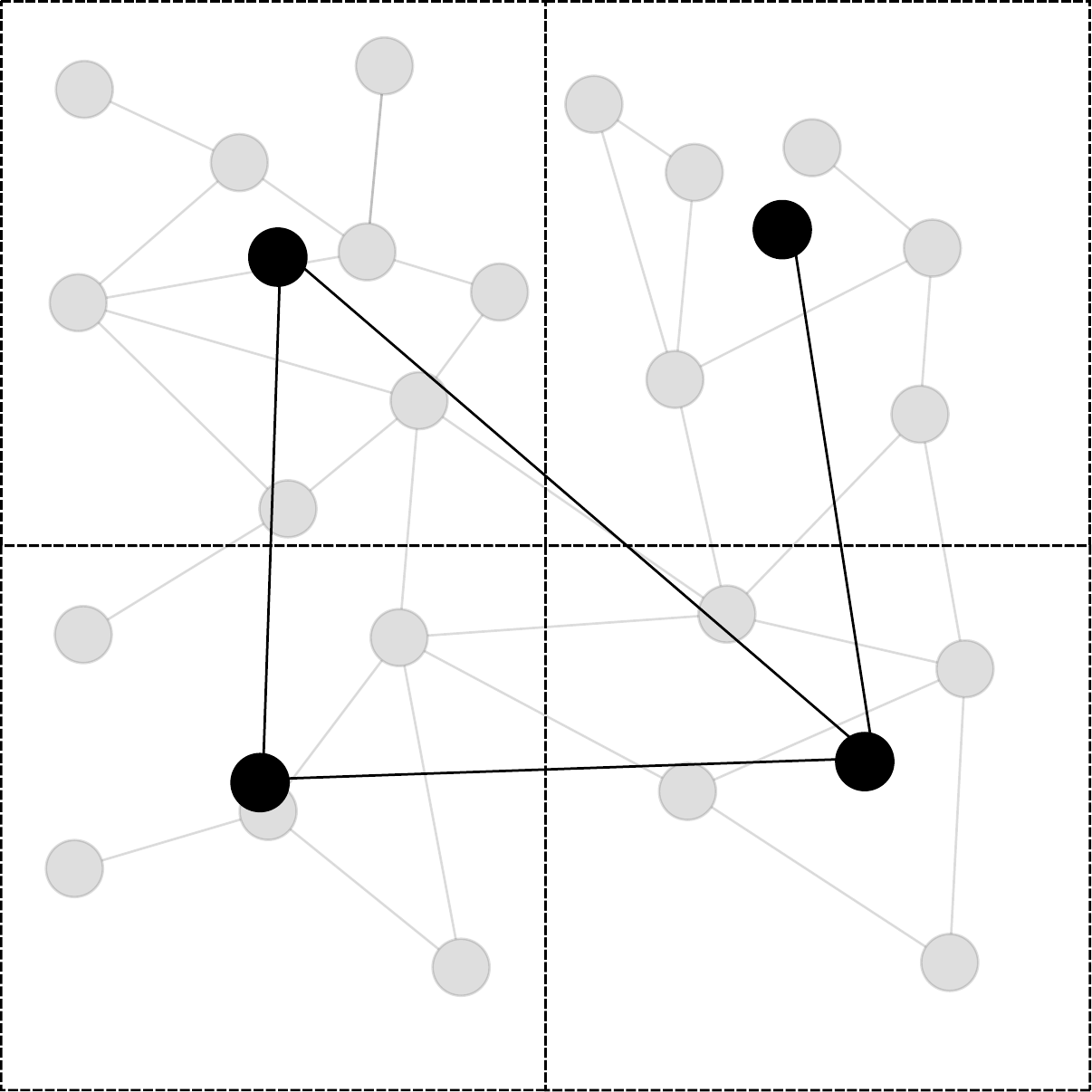}
\caption{Network clustering algorithm: The original nodes inside a clustering cell are aggregated into one representative node. 
}
\label{fig:clustering}
\end{figure}

\begin{figure}
\includegraphics[scale=0.5]{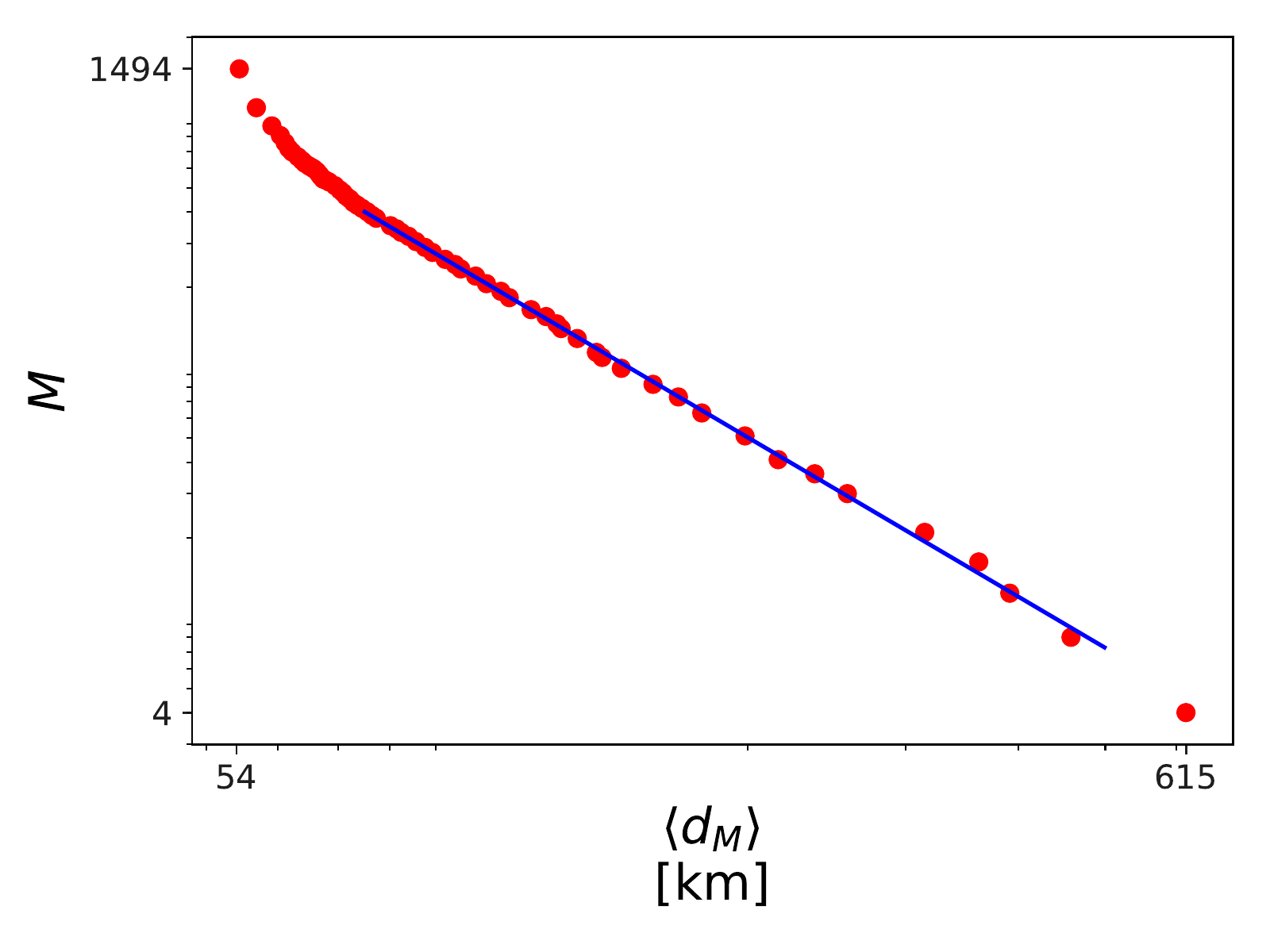}
\caption{Relation between the network size $M$ of the coarse-grained representation of the EU electricity grid resulting from the clustering algorithm with increasing cell areas, and the respective average link length $\langle d_{M}\rangle$. The network size decreases from $1494$ nodes (the original network) to $4$ nodes, with the average link length increasing from $\approx 54$ km to $\approx 615$ km. The blue link shows a fit to $\langle d_{M}\rangle \propto M^{\gamma}$ in the range $\langle d_{M}\rangle \in [75,500]$ km, yielding $\gamma\approx -2.11$.}
\label{fig:size_length}
\end{figure}

\begin{figure}
\includegraphics[scale=0.5]{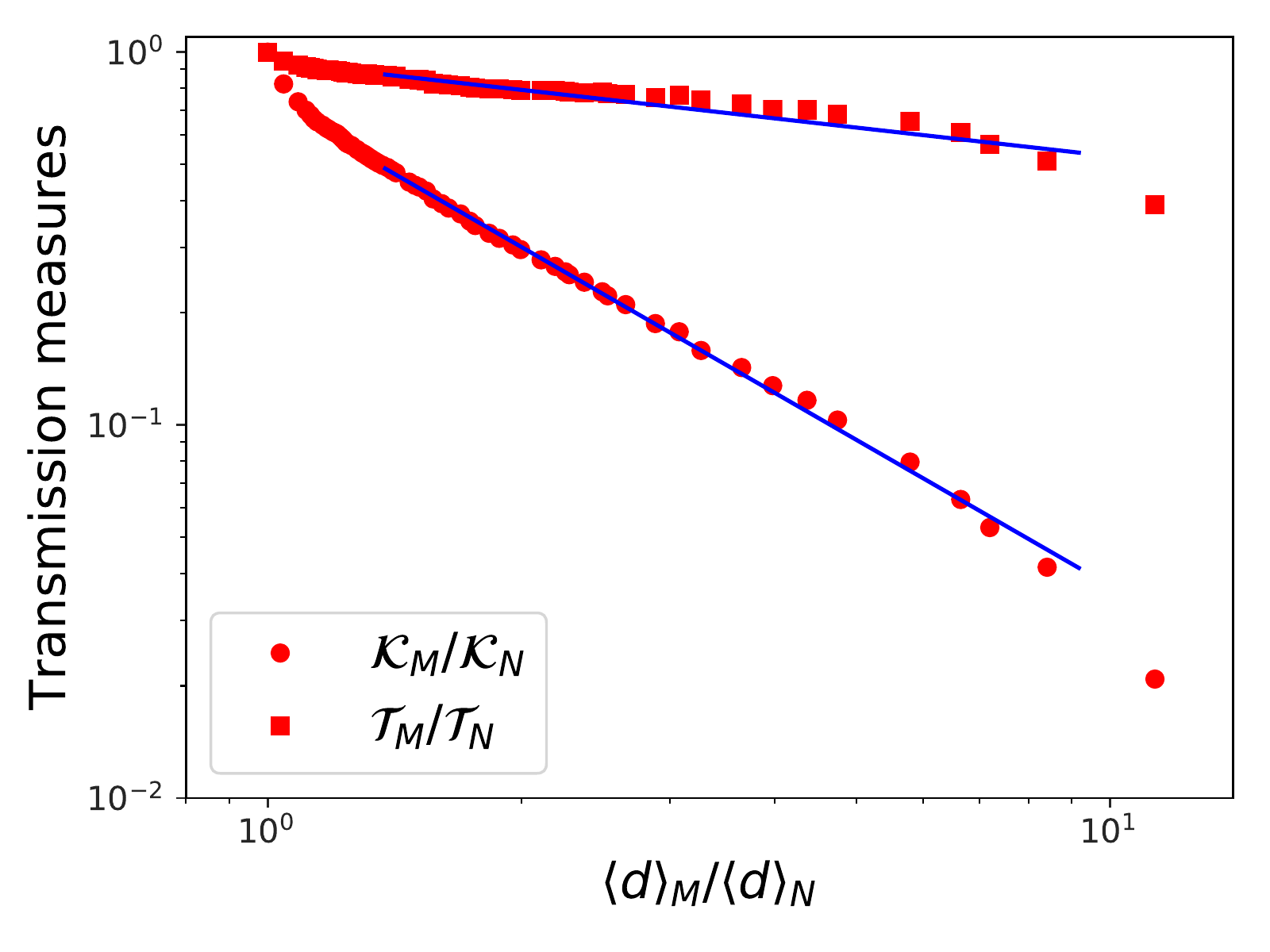}
\caption{Relative total transmission capacity $\mathcal{K}_{M}/\mathcal{K}_{N}$ and cost $\mathcal{T}_{M}/\mathcal{T}_{N}$ for coarse-grained representations of the simplified EU electricity system model with network size $M$ and relative spatial scale $\langle d_{M}\rangle/\langle d_{N}\rangle$. The blue lines show simple fits to power laws for both expressions in the range $\langle d_{M}\rangle\in[75,500]$ km, yielding approximate scaling exponents of $\eta_{1}\approx -1.30$ for the total transmission capacity, and $\eta_{2}\approx - 0.25$ for the total transmission cost.}
\label{fig:eu_scaling}
\end{figure}

We are interested in the scaling properties of both the total transmission capacity $\mathcal{K}$ and cost $\mathcal{T}$ under network coarse-graining, that is for representations of the original system with different spatial resolutions. The coarse-graining is realised by applying a simple spatial clustering procedure to the model of the European transmission grid. We overlay a two-dimensional lattice containing $M$ non-empty aggregation cells of equal area on top of the original network with $N$ nodes. The nodes contained in each cell are replaced by one representative node, located at the average position of the aggregated nodes in the cell. Two coarse-grained nodes are connected by a link in the newly created network, if there is at least one link between the respective sets of underlying nodes in the original network (see fig.~\ref{fig:clustering}).
By successively increasing the size of the clustering cells, we obtain aggregated networks with sizes from the original $N=1494$ nodes down to $M=4$ nodes (see fig.~\ref{fig:eu_clustering}). The spatial scale of each system is expressed by the average link length $\langle d_{M}\rangle$. Figure~\ref{fig:size_length} shows that the relation between the coarse-grained network size $M$ and corresponding average link length$\langle d_{M} \rangle$ for the the range $75$ km to $500$ km can be expressed as $M\propto \langle d_{M}\rangle^{\gamma}$ with $\gamma \approx -2.11$, compared to $\gamma=-2$ which would hold for a two-dimensional lattice. 
For each coarse-grained network, the nodal mismatch time series are determined by summation of the time series of all original nodes in the corresponding aggregation cells. Applying the same power flow equations as for the original network, we obtain the flow statistics and the resulting infrastructure measures $\mathcal{K}_{M}$ and $\mathcal{T}_{M}$ for the coarse-grained network with size $M$ and spatial scale $\langle d_{M}\rangle$. Figure~\ref{fig:eu_scaling} shows $\mathcal{K}_{M}/\mathcal{K}_{N}$ and $\mathcal{T}_{M}/\mathcal{T}_{N}$ as a function of $\langle d_{M}\rangle/\langle d_{N}\rangle$. We observe that the results for both measures decrease under coarse-graining. Applying a simple fit to a power law
\begin{equation}
\frac{\mathcal{K}_{M}}{\mathcal{K}_{N}} \propto \left(\frac{\langle d_{M}\rangle}{\langle d_{N}\rangle
}\right)^{\eta_1}\quad , \quad
\frac{\mathcal{T}_{M}}{\mathcal{T}_{N}} \propto \left(\frac{\langle d_{M}\rangle}{\langle d_{N}\rangle
}\right)^{\eta_2}~,
\end{equation}
for the range $\langle d_{M}\rangle\in[75,500]$ km we observe scaling exponents $\eta_{1}\approx -1.30$ and $\eta_{2}\approx - 0.25$. This shows in particular that the transmission costs of the system only scale weakly with the spatial resolution of the network representation.
\section{Analytical approximations for transmission capacity measures}
\begin{figure}
\includegraphics[scale=0.46]{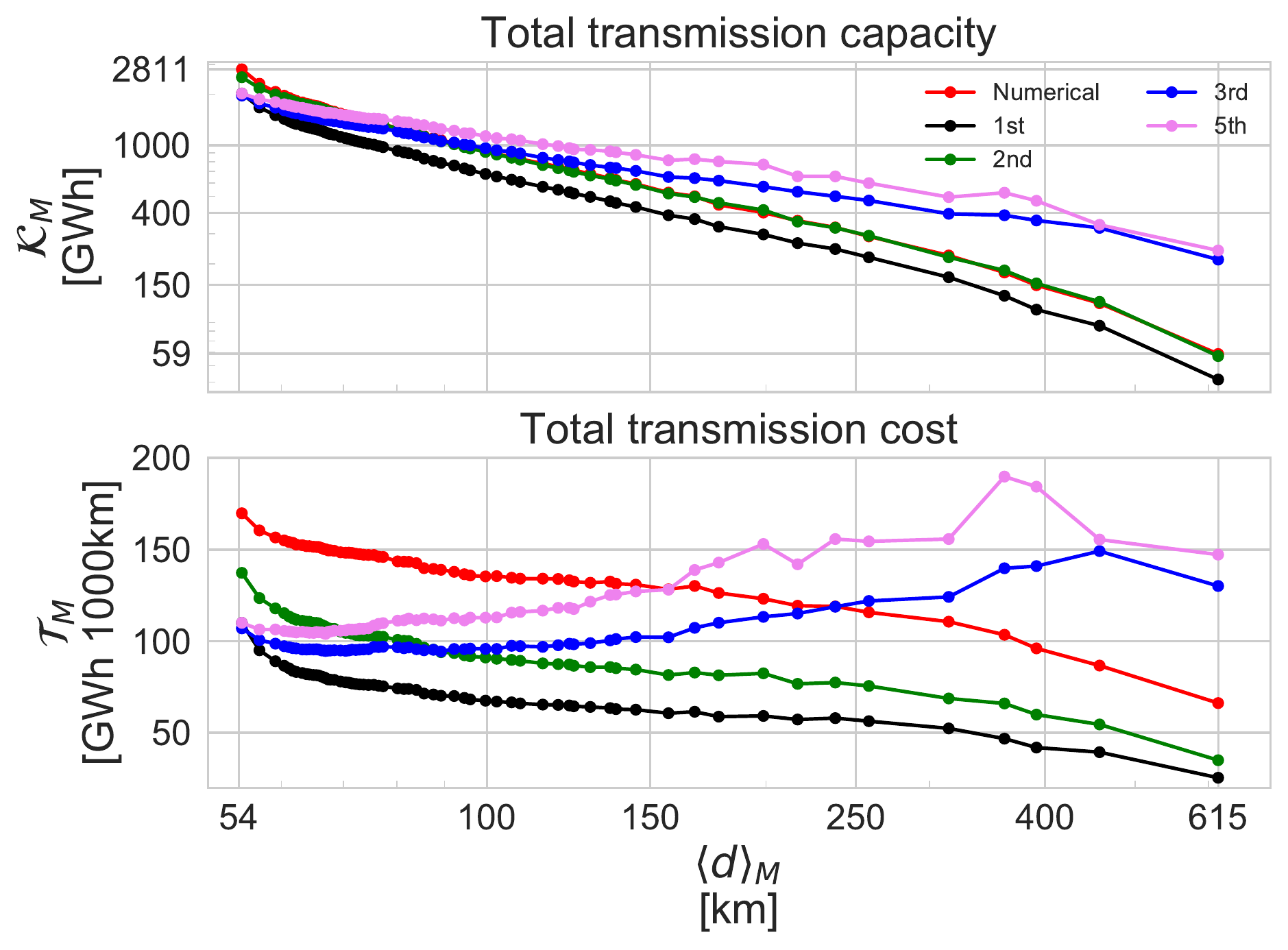}
\caption{Full numerical results (red line) and different approximations for the total transmission capacity (top) and cost (bottom) in a simplified model of a highly renewable EU electricity system under spatial coarse-graining. The spatial scale of the system is expressed by the average link length $\langle d_{M}\rangle$. The different approximations refer to the following equations. 1st (black): eqs.~(\ref{eq:first1}) and ~(\ref{eq:first2}). 2nd (green): eq.~(\ref{eq:second}). 3rd (blue): eq.~(\ref{eq:third}). 5th (violet): eq.~(\ref{eq:final}). The fourth approximation in eq.~(\ref{eq:fourth}) yields very similar results compared to the fifth approximation in eq.~(\ref{eq:final}) and is not depicted here. For the transmission capacity, over a wide range the red line showing the full numerical results is covered by the green line representing the results obtained after implementing the second approximation in eq.~(\ref{eq:second}). Note that the scale for the y-axis is chosen to be logarithmic for the transmission capacity, whereas it is linear for the transmission cost.}
\label{fig:eu_approximations}
\end{figure}

According to eq.~(\ref{eq:gauge2}), the power flows depend on the mismatch statistics $\Delta_{n}(t)$ at the individual nodes $n$ and on the network topology expressed in the PTDF matrix $H_{ln}$. An exact analytical description of the scaling of $\mathcal{K}$ and $\mathcal{T}$ under coarse-graining as observed in the last section in general is prevented by the correlated structure of the mismatch pattern $\Delta_{n}(t)$, the changes of the grid topology under clustering, and the difficulties arising from the consideration of tails of the flow distribution in the definition in eq.~(\ref{eq:capacity}). In the following we will introduce a series of subsequent  approximations for $\mathcal{K}$ and $\mathcal{T}$ leading to simplified expressions for these transmission infrastructure measures, which allow to analytically estimate the respective scaling properties under coarse graining.\\
As a first step we approximate the original mismatch distribution using a multivariate normal distribution with mean $\bar{\mathbf{\Delta}}$ and covariance matrix $\mathbf{C}^{\Delta}$ with $C_{mn}^{\Delta}=\text{Cov}(\Delta_{m},\Delta_{n})$. Here $\bar{\mathbf{\Delta}}$ denotes the average mismatch vector with entries $\langle\Delta_{n}\rangle$.  From the linearity of the power flow equations it follows that the resulting flow distribution on the links itself is a multivariate normal distribution with mean flow vector $\bar{\mathbf{F}}=\mathbf{H}\bar{\mathbf{\Delta}}$ and flow covariance matrix $\mathbf{C}^{F}=\mathbf{H}\mathbf{C}^{\Delta}\mathbf{H}^{T}$. In particular, in this case the quantile in eq.~(\ref{eq:capacity}) can be expressed using the error function $\text{erf}(\cdot)$:
\begin{equation}
q = \frac{1}{2}\left\{
\text{erf}\left(\frac{\mathcal{K}_{l} - \langle F_{l} \rangle}{\sqrt{2C^{F}_{ll}}}\right)
+
\text{erf}\left(\frac{\mathcal{K}_{l} + \langle F_{l} \rangle}{\sqrt{2C^{F}_{ll}}}\right)
\right\}~.
\end{equation}
Despite the heterogeneous solar and wind generation layouts inside the countries of the given EU electricity system model, the transmission capacities are dominated by the distribution of power flows resulting from the fluctuations in the underlying mismatch pattern, rather than by the average flows resulting from heterogeneities in the distribution of the average mismatches. Consequently we can assume $\mathcal{K}_{l} \gg \langle F_{l} \rangle$ and approximate $q=\text{erf}\left(\mathcal{K}_{l}/\sqrt{2C^{F}_{ll}}\right)$. We invert this relation and obtain the expression
\begin{equation}
\mathcal{K}_{l} = \sqrt{2}\text{erf}^{-1}(q)\sigma(F_{l})~,
\end{equation}
where we have used $\sigma(F_{l})=\sqrt{C_{ll}^{F}}$, which denotes the standard deviation of the flow distribution $p(F_{l})$. The total capacity and transmission cost then read
\begin{align}
\label{eq:first1}
\mathcal{K} &=\sqrt{2}\text{erf}^{-1}(q)\left(\sum_{l}\sigma(F_{l})\right) = 
\sqrt{2}\text{erf}^{-1}(q)L\langle \sigma(F_{l}) \rangle~,\\
\label{eq:first2}
\mathcal{T} &=\sqrt{2}\text{erf}^{-1}(q)\left(\sum_{l}d_{l}\sigma(F_{l})\right) = 
\sqrt{2}\text{erf}^{-1}(q)L\langle d_{l}\sigma(F_{l}) \rangle~,
\end{align}
with the average taken over all $L$ links of the network, respectively.\\
The red curve in fig.~\ref{fig:eu_approximations} shows the numerical results for $\mathcal{K}_{M}$ and $\mathcal{T}_{M}$ as defined in eqs.~(\ref{eq:capacity}) and~(\ref{eq:cost}). The results according to the first approximation in eqs.~(\ref{eq:first1}) and~(\ref{eq:first2}) are depicted by the black curve. We observe that for both the transmission capacity and cost this approximation yields smaller values, thus underestimating the respective infrastructure needs. This can be explained by tails in the mismatch distributions $\Delta_{n}(t)$. Replacing the original distribution by a multivariate normal distribution reduces these tails, which in turn reduces the tails and thus higher quantiles of the power flow distributions $F_{l}(t)$, which by definition reduces the capacity and cost measures.\\
It turns out that for a further analytical treatment it is advantageous to work with the average variance of the flow distribution $\langle \sigma^2(F_{l})\rangle$ instead of the average standard deviation $\langle \sigma(F_{l})\rangle$. We thus substitute
\begin{align}
\label{eq:homo1}
\langle \sigma(F_{l}) \rangle \approx \sqrt{\langle \sigma^2(F_{l})\rangle}~,\\
\label{eq:homo2}
\langle d_{l} \sigma(F_{l}) \rangle \approx \langle d_{l} \rangle \sqrt{\langle \sigma^2(F_{l})\rangle}~.
\end{align}
Due to $\langle \sigma(F_{l}) \rangle^2\leq \langle \sigma^2(F_{l})\rangle$, for the simplified EU model this substitution will increase the transmission capacity measure $\mathcal{K}$. For the transmission cost measure $\mathcal{T}$ correlations between $d_{l}$ and $\sigma(F_{l})$ have to be taken into account, but in general also this measure will increase under the approximation. Using eqs.~(\ref{eq:homo1}) and~(\ref{eq:homo2}), the total transmission capacity $\mathcal{K}$ and cost $\mathcal{T}$ can be written as follows:
\begin{align}
\mathcal{K} &= \frac{1}{\langle d \rangle} \mathcal{T}=
\sqrt{2}\text{erf}^{-1}(q)L\sqrt{\frac{\sum_{l}\sigma^2(F_{l})}{L}} \nonumber \\
&=
\sqrt{2}\text{erf}^{-1}(q)\sqrt{L}\sqrt{\text{Tr}\left[\mathbf{C}^{F}\right]} \nonumber \\
\label{eq:second}
&=
\sqrt{2}\text{erf}^{-1}(q)\sqrt{L}\sqrt{\text{Tr}\left[\mathbf{H}\mathbf{C}^{\Delta}\mathbf{H}^{T}\right]}~.
\end{align}
The green curve in fig.~\ref{fig:eu_approximations} shows the resulting approximated value for the transmission infrastructure measures $\mathcal{K}_{M}$ and $\mathcal{T}_{M}$ according to eq.~(\ref{eq:second}). These results are denoted as the second approximation. We can see that the curves are as expected shifted to larger values compared to the ones obtained for the first approximation in eqs.~(\ref{eq:first1}) and~(\ref{eq:first2}). In particular, for the specific system under study the errors due to both approximations almost cancel each other for the total capacity $\mathcal{K}_{M}$, leading to a result which is close to the original numerical value.\\
Spatio-temporal correlations in both the load and renewable generation time series translate into correlations in the mismatch time series $\Delta_{n}(t)$. Neglecting these correlations allows to further simplify the expressions for the infrastructure measures $\mathcal{K}$ and $\mathcal{T}$. Approximating in that case $\mathbf{C}^{\Delta}\approx \text{diag}(\sigma^2(\Delta_{n}))$, we diagonalise the real symmetric matrix $\mathbf{H}^{T}\mathbf{H}$ with eigenvalues $\mu_{k}$ and obtain
\begin{align}
\label{eq:third}
\mathcal{K}&=\frac{1}{\langle d \rangle} \mathcal{T}\nonumber\\
&\approx \sqrt{2}\text{erf}^{-1}(q)\sqrt{L}
\sqrt{
\sum_{k}\mu_{k}\left(\sum_{n}\left(u_{n}^{(k)}\right)^2\sigma^2(\Delta_{n})\right)
}~,
\end{align}
where $u_{n}^{(k)}$ denotes the $n$th component of the $k$th eigenvector of $\mathbf{H}^{T}\mathbf{H}$. In fig.~\ref{fig:eu_approximations} the expressions in eq.~(\ref{eq:third}), denoted as the third approximation, are shown as a blue curve for both the total capacity $\mathcal{K}$ and cost $\mathcal{T}$. We observe that compared to the second approximation, for the original system and spatial scales up to $\langle d_{M}\rangle \approx 90$ km this simplification reduces the results for both transmission infrastructure measures, wheareas for the more coarse-grained systems with $\langle d_{M} \rangle > 90$ km the situation is reversed. Considering the original system, we expect correlations to increase the infrastructure needs, because geographically close regions will show similar weather and thus in particular wind generation patterns, leading to power transmission over longer distances, resulting in higher fluctuating power flows. For coarse-grained networks, these shorter-range spatial correlations in the renewable generation still increase the flow fluctuations by being incorporated in the mismatch variances $\sigma^{2}(\Delta_{n})$ for the aggregated nodes. Nevertheless, although decreasing, spatial correlations are also present over larger distances, for instances due to similar load patterns, the day and night cycle in solar generation, or large scale weather patterns. Neglecting these correlations represented in the off-diagonal elements of $\mathbf{C}^{\Delta}$ overestimates the heterogeneity of the system, which leads to increasing transmission infrastructure needs as depicted in fig.~\ref{fig:eu_approximations} for $\langle d_{M}\rangle>90$ km.\\
In order to further simplify the expressions for the transmission infrastructure measures $\mathcal{K}$ and $\mathcal{T}$, we substitute $\sigma^2(\Delta_{n})\approx \langle \sigma^2(\Delta_{n})\rangle$, which yields the fourth approximation for the transmission infrastructure measures
\begin{align}
\mathcal{K} &= \frac{1}{\langle d \rangle} \mathcal{T}\nonumber\approx
\sqrt{2}\text{erf}^{-1}(q)\sqrt{
L\langle \sigma^2(\Delta_{n})\rangle
}\sqrt{
\text{Tr}\left[\mathbf{H}\mathbf{H}^{T}\right]
}\\
\label{eq:fourth}
&= 
\sqrt{2}\text{erf}^{-1}(q)\sqrt{
L\langle \sigma^2(\Delta_{n})\rangle
}\sqrt{\sum_{k}\mu_{k}}~.
\end{align}
Recall that the PTDF matrix has been gauged according to eqs.~(\ref{eq:gauge2}) and (\ref{eq:gauge_c}) to incorporate the balancing. For a uniform balancing, that is in our case $\langle L_{n} \rangle = (\sum_{k}\langle L_{k} \rangle)/N$, it can be shown that for $\mathbf{H}=\mathbf{K}^{T}\mathbf{L}^{\dagger}$ the correct balancing is already incorporated and we obtain $\hat{c}_{l}=0$. If we apply this approximation of uniform balancing, the expression $\mathbf{H}^{T}\mathbf{H}$ simplifies to
\begin{equation}
\mathbf{H}^{T}\mathbf{H}=\mathbf{L}^{\dagger}\mathbf{K}\mathbf{K}^{T}\mathbf{L}^{\dagger}=
\mathbf{L}^{\dagger}\mathbf{L}\mathbf{L}^{\dagger}=\mathbf{L}^{\dagger}~,
\end{equation}
where we have used $\mathbf{K}\mathbf{K}^{T}=\mathbf{L}$. The sum over the eigenvalues of $\mathbf{H}\mathbf{H}^{T}$ in eq.~(\ref{eq:fourth}) is thus given by the sum over the eigenvalues of $\mathbf{L}^{\dagger}$, which correspond to the inverses of the non-zero eigenvalues of the network Laplacian $\mathbf{L}$. This sum is proportional to the so-called Kirchhoff index or Quasi-Wiener index $\text{Kf}$ of the network~\cite{gutman1996,klein1993}:
\begin{equation}
\text{Kf} = N \text{Tr}\left[\mathbf{L}^{\dagger}\right] = N\sum_{k=1}^{N-1}\frac{1}{\lambda_{k}}~.
\end{equation}
The eigenvalues $\lambda_{k}$ in this relation are ordered in descending order, such that $\lambda_{N}=0$. The Kirchhoff index denotes the sum of resistance distances between all pairs of vertices in the network. Here the resistance distance is given by the resistance between two nodes in a corresponding network, in which all individual links have unit resistance~\cite{klein1993}. Incorporating all simplifications presented in this section we  can write for the total transmission capacity and cost of the system the following final fifth approximation:
\begin{equation}
\label{eq:final}
\mathcal{K} = \frac{1}{\langle d \rangle}\mathcal{T}=
\text{erf}^{-1}(q)\sqrt{\langle k \rangle \text{Kf}}\sqrt{\langle \sigma^2(\Delta)\rangle}~.
\end{equation}
In this relation we have used $2L=\langle k \rangle N$, with $\langle k \rangle$ denoting the average degree of the network. It is appealing that in eq.~(\ref{eq:final}) the influence from the nodal mismatch statistics and the role of the network topology are separated. In fig.~\ref{fig:eu_approximations} we display this final result, denoted as the fifth approximation, by the violet curve. The previous fourth approximation in eq.~(\ref{eq:fourth}) yields very similar results and is not depicted in these figures. In conclusion we observe that while in particular due to the non-consideration of correlations the details of the relations between the transmission infrastructure measures and the spatial scale are not represented by the expressions in eq.~(\ref{eq:final}), the general trend is well approximated.
\section{Scaling of transmission capacities and costs under network aggregation}
\begin{figure}
\includegraphics[scale=0.43]{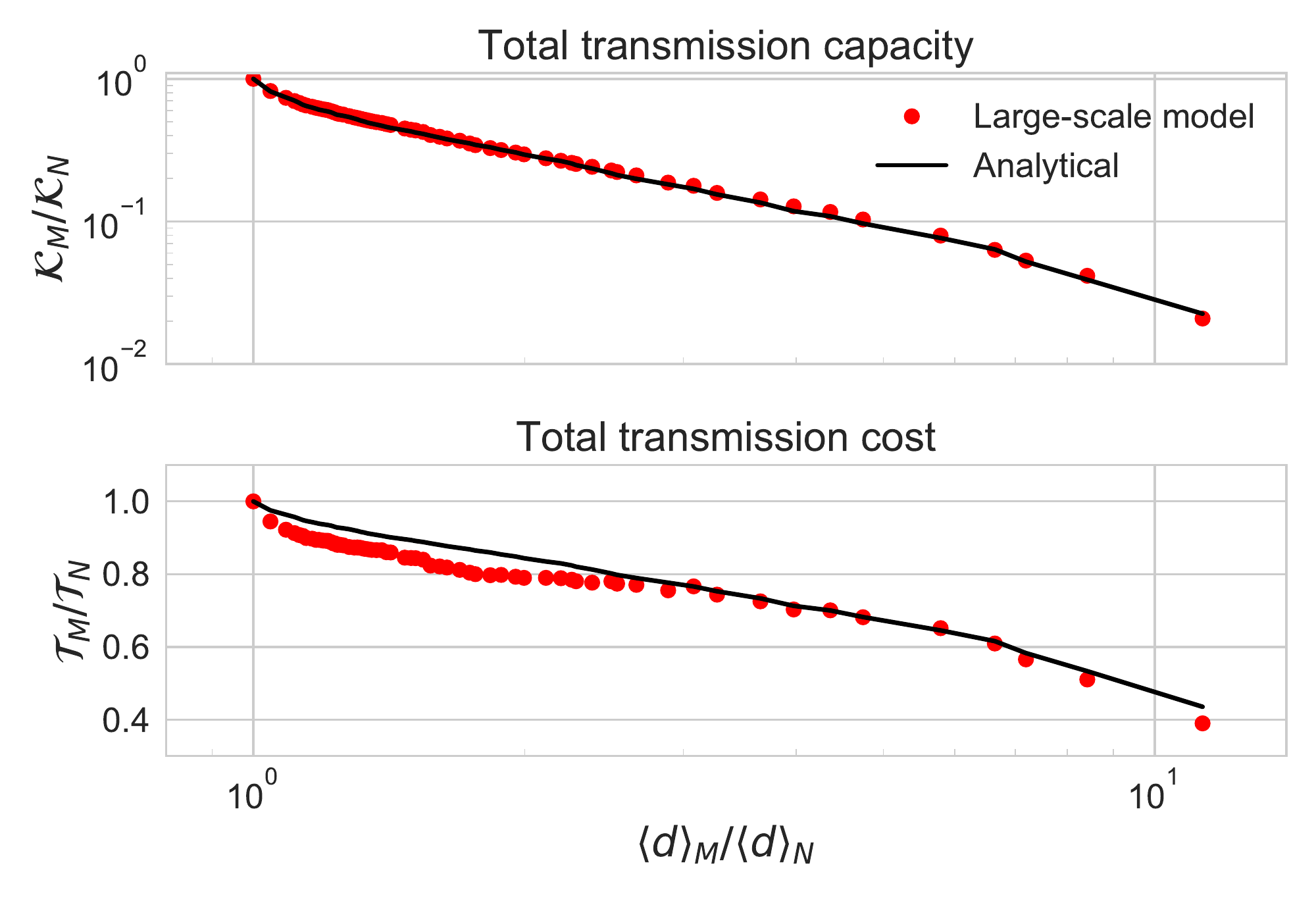}
\caption{Scaling of total transmission capacity (top) and cost (bottom) for the EU electricity system model. The numerical results (red dots) are compared to the analytical estimate (black line) according to eqs.~(\ref{eq:scaling1}) and (\ref{eq:scaling2}). Note that the scale for the y-axis is chosen to be logarithmic for the transmission capacity, whereas it is linear for the transmission cost.}
\label{fig:eu_scaling_analytical}
\end{figure}

How does the simplified expression for the transmission infrastructure measures $\mathcal{K}$ and $\mathcal{T}$ in eq.~(\ref{eq:final}) changes under coarse-graining? Due to the summation of nodal time series inside an aggregation cell we can assume as a first approximation 
\begin{equation}
\langle \sigma^2(\Delta)\rangle _{M} \approx \frac{N}{M} \langle \sigma^2(\Delta) \rangle_{N}~,
\end{equation}
with $N/M$ the average number of original nodes aggregated inside one cell, and the index $N$ and $M$ referring to the observable evaluated for the original network with $N$ nodes, or the coarse-grained network with $M$ nodes. For spatial infrastructure networks the degree distribution is often very homogeneous~\cite{barthelemy2011}, which suggests to approximate a constant average degree $\langle k \rangle_{N} \approx \langle k \rangle_{M}$ in eq.~(\ref{eq:final}) for different spatial resolutions of the network. In order to obtain an analytical estimate of the scaling properties of the Kirchhoff index $\text{Kf}$, we approximate the original and each coarse-grained network by a two-dimensional lattice with approximately the same number of nodes, respectively. The Laplacian eigenvalues of a 2D lattice graph with $N=\sqrt{N}\times \sqrt{N}$ nodes are given as~\cite{vanmieghem2010}:
\begin{equation}
\lambda_{n,m} = 4\sin^2\left(\frac{\pi n}{2\sqrt{N}}\right) + 4 \sin^2\left(\frac{\pi m}{2\sqrt{N}}\right)~.
\end{equation}
Here $0\leq n,m < \sqrt{N}$, with $\lambda_{0,0}=0$. The Kirchhoff index corresponding to the sum over the inverse non-zero eigenvalues for large $N$ then can be approximated as follows:
\begin{align}
\text{Kf} &= N \sum_{\substack{0\leq n,m < \sqrt{N} \\ n=m\neq 0}}\frac{1}{\lambda_{n}}\\
&\approx \frac{N^2}{\pi^2}\sum_{\substack{0\leq n,m < \sqrt{N} \\ n=m\neq 0}}\frac{1}{
n^2 + m^2}\\
&\approx
\frac{N^2}{\pi^2}\int_{0}^{\frac{\pi}{2}} \int_{\frac{1}{\sqrt{N}}}^{1}\frac{r\:d \phi\:d r}{r^2}\\
&=\frac{N^2 \ln \left[\sqrt{N}\right]}{2\pi} = \frac{N^2 \ln N}{4\pi}~.
\end{align}
Under the spatial clustering procedure from $N$ original nodes to $M$ aggregated nodes, the scaling of the Kirchhoff index thus can be approximated as
\begin{equation}
\text{Kf}_{M} \approx \frac{M^2 \ln M}{N^2 \ln N}\text{Kf}_{N}~.
\end{equation}
Approximating the area covered by the original network as $N\langle d \rangle_{N}^2$, we can write $N \langle d \rangle_{N} ^2 \approx M \langle d\rangle_{M}^2$, which is a first-order approximation to $M\propto\langle d_{M}\rangle^{-2.11}$ found numerically for the EU electricity system model. Collecting all relations discussed in this section we finally obtain
\begin{align}
\label{eq:scaling1}
\mathcal{K}_{M} &\approx \sqrt{\frac{M \ln M}{N \ln N}}\mathcal{K}_{N} \approx \left(\frac{\langle d \rangle_{M}}{\langle d \rangle_{N}}\right)^{-1}\sqrt{1-\frac{2\ln\left[\frac{\langle d \rangle_{M}}{\langle d \rangle_{N}}\right]}{\ln N}}\:\mathcal{K}_{N}~,\\
\label{eq:scaling2}
\mathcal{T}_{M} &\approx \sqrt{\frac{\ln M}{\ln N}}\mathcal{T}_{N} \approx \sqrt{1-\frac{2\ln\left[\frac{\langle d \rangle_{M}}{\langle d \rangle_{N}}\right]}{\ln N}}\:\mathcal{T}_{N}~.
\end{align}
Figure~\ref{fig:eu_scaling_analytical} compares the scaling of $\mathcal{K}$ and $\mathcal{T}$ dependent on the length scale $\langle d \rangle_{M}$ for the simplified renewable EU electricity system model with the analytical result in eqs.~(\ref{eq:scaling1}) and~(\ref{eq:scaling2}), respectively.
The figure shows that for this system the analytical scaling provides an accurate description of the numerical results. In particular, by considering the relative measures $\mathcal{K}_{M}/\mathcal{K}_{N}$ and $\mathcal{T}_{M}/\mathcal{T}_{N}$, the impact of systemic errors due to the approximated calculation of transmission capacities and costs have been reduced. Despite the non-grid like topology and the neglecting of correlations in the mismatch data, the essential scaling properties of transmission infrastructure measures under coarse-graining in the model of the European power grid are thus described by the analytical relations in eqs.~(\ref{eq:scaling1}) and~(\ref{eq:scaling2}). It should be emphasized that in this context the transmission capacity cost only scales weakly with the spatial resolution of the system.
\section{Conclusion}
Due to the complexity of the power grid, models of the electricity system often consider a coarse-grained network representation in which the topology, load and generation patterns below a given spatial scale are aggregated into representatives nodes. Given that depending on the model the spatial scale might range from resolving single transmission stations to countries as network nodes, it is important to understand how the resulting infrastructure objectives depend on this coarse-graining procedure~\cite{hoersch2017}. In this contribution, we study scaling properties of transmission infrastructure measures under coarse-graining in a simplified, but data-driven and spatially detailed model of a highly renewable European electricity system. We observe that the transmission capacity cost only scales weakly with the spatial scale of the system. By applying a series of approximations we obtain an analytical description of the scaling properties under coarse-graining, which for the model system describes the numerical results remarkably accurately.\\
The results presented in this article suggest future research in different directions. With respect to models of the electricity system, it would be interesting to study to what extent the analytical description still holds for more heterogeneous layouts, in particular concerning the distribution of renewable generation capacities~\cite{eriksen2017}. In the present article we defined transmission capacities in retrospect, determined by the flow statistics at the different levels of coarse-graining. Alternatively one could also obtain macroscopic power flows by aggregating microscopic flows~\cite{poulsen2016}, or directly aggregate the existing transmission infrastructure without calculating power flows~\cite{hoersch2017}. Beyond the field of electricity system modelling, the role of coarse-graining for the determination of system objectives is also relevant for other infrastructure and transport networks~\cite{barthelemy2011}. From a complex networks perspective, investigating the scaling of transport properties under network clustering for generic network models, for instance geometric, small-world or scale-free networks~\cite{newman2010}, would allow to further understand the relation between network topology, nodal dynamics, and emerging flow patterns on different spatial scales.
\begin{acknowledgments}
Mirko Sch\"afer is funded by the Carlsberg Foundation Distinguished Postdoctoral Fellowship. We thank Jonas H\"{o}rsch and Tue Jensen for fruitful discussions.
\end{acknowledgments}


\bibliographystyle{unsrt}
\bibliography{references}


\end{document}